\documentclass[aps,prl,twocolumn,showpacs,preprintnumbers,groupedaddress]{revtex4}
\usepackage{graphicx} 

\begin{document}


\title{
Gain in a quantum wire laser of high uniformity 
}

\author{
Hidefumi Akiyama${}^{1,2,3}$, Loren N. Pfeiffer${}^2$, Masahiro Yoshita${}^1$, Aron Pinczuk${}^{2,3}$, Peter B. Littlewood${}^{2,4}$, Ken W. West${}^2$, Manyalibo J. Matthews${}^2$, and James Wynn${}^2$
}

\affiliation{
${}^1$Institute for Solid State Physics, University of Tokyo, \\ 
5-1-5 Kashiwanoha, Kashiwa, Chiba 277-8581, Japan\\ 
${}^2$Bell Laboratories, Lucent Technologies, \\ 
600 Mountain Avenue, Murray Hill, NJ 07974, USA \\
${}^3$Department of Physics, Columbia University, \\ 
New York, NY 10027, USA \\
${}^4$TCM group, Cavendish Laboratory, University of Cambridge, 
Cambridge CB3 0HE, UK
}

\date{\today}

\begin{abstract} 
A multi-quantum wire laser operating in the 1-D ground state has been achieved in a very high uniformity structure that shows free exciton emission with unprecedented narrow width and low lasing threshold.  Under optical pumping the spontaneous emission evolves from a sharp free exciton peak to a red-shifted broad band.  The lasing photon energy occurs about 5 meV below the free exciton.  The observed shift excludes free excitons in lasing and our results show that Coulomb interactions in the 1-D electron-hole system shift the spontaneous emission and play significant roles in laser gain. 
\end{abstract}

\maketitle

Quantum wire lasers provoke fundamental questions stemming from the singular nature of one-dimensional (1-D) densities of state \cite{ArakawaAPL,KaponPRL,WegscheiderPRL,OgawaPRLBennerELHuPRL,GlutschBrinkmannWalck,SzymanskaPRB,RossiPRBTassonePRLDassarmaPRL}. A quantum wire laser was first achieved by Kapon and coworkers in 1989 \cite{KaponPRL}, though lasing occurred only at higher subbands in multi-mode wires. In 1993, Wegscheider and coworkers \cite{WegscheiderPRL} demonstrated ground-state lasing in quantum wires. They found that the lasing energy was exactly at the peak of excitonic spontaneous emission, and was nearly independent of pump levels. This suggested absence of band-gap renormalization and an enhanced stability of 1-D excitons. Therefore, the origin of gain was ascribed to excitons. 

As for the enhanced stability and activity of 1-D excitons in narrow quantum wires, there have been numbers of theories \cite{OgawaPRLBennerELHuPRL,GlutschBrinkmannWalck,SzymanskaPRB,RossiPRBTassonePRLDassarmaPRL} and experiments \cite{AkiyamaJP}, for example, on binding energy \cite{SomeyaPRL}, oscillator strength \cite{AkiyamaPRB96b}, and many-body effects \cite{RubioSSC,AmbigapathyPRL,SiriguPRB,AkiyamaSSC}. In particular, lack of red-shift in photoluminescence (PL) under high photo-excitation levels has been reported in various 1-D systems \cite{WegscheiderPRL,RubioSSC,AmbigapathyPRL,SiriguPRB,AkiyamaSSC}.

General arguments suggest that free excitons cannot cause lasing \cite{private} since the electron-hole population will not become inverted until a density approaching one exciton per Bohr radius. At such a density the excitonic correlations will be transient and such a system is probably better described as a Coulomb-correlated electron-hole plasma. The argument can be subverted if the excitons become localized \cite{SiriguPRB} in low energy states produced by disorder or impurities, and therefore it is important to study systems of extremely high uniformity.

\begin{figure}
\includegraphics[width=.3\textwidth]{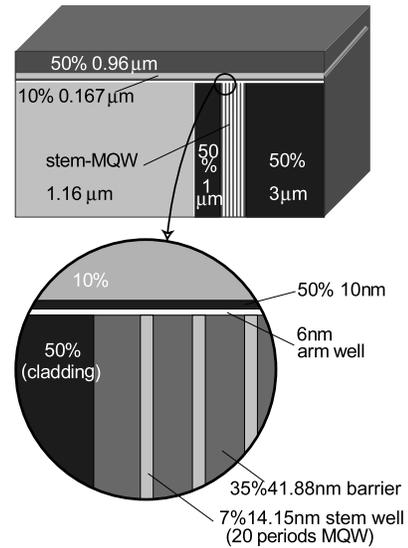}
\caption{
Schematic of a T-wire laser structure. Percentages show Al-concentration $x$ in Al${}_x$Ga${}_{1-x}$As. The laser contains 20 periods of T-wires defined by 7\%-Al filled 14.15 nm stem wells and 6 nm GaAs arm well embedded in a T-shaped optical waveguide with a 500 $\mu$m cavity between uncoated facets. Point PL spectroscopy is performed from the top surface. Lasing experiment is achieved by optical pumping in a filament shape via the top surface, where stimulated and spontaneous emissions are detected via the front and top surfaces. 
}
\label{1}
\end{figure}

In this letter, we report stimulated and spontaneous emissions of a quantum wire laser with greatly improved uniformity. Remarkably, optically pumped lasing is observed with a lower threshold than in previous quantum wire lasers \cite{WegscheiderPRL,RubioSSC}. Free exciton emission is extraordinarily sharp for quantum wires, while localized exciton emission is very weak and does not seem to play an important role in the gain of this laser. Lower lasing threshold and enhanced free exciton properties are attributed to uniformity and greatly reduced interface roughness. We find that near threshold lasing emission occurs at 5 meV below the free exciton. This is a result that demonstrates that lasing gain in the T-wire structure is not linked to the free exciton emission. At high pumping levels, well above threshold for lasing, the free exciton peak is quenched in spontaneous emission while a new broad emission band at lower energy becomes dominant. The red-shift and broadening of the emission suggest formation of a 1-D electron-hole plasma. The weak dependence of the energy of the emission band on photo-excitation intensity at high pumping levels indicates that electron-hole Coulomb correlations remain strong in the plasma. The observed laser photon energies, being slightly below the peak energy of the spontaneous emission, are evidence that the gain for lasing is created by the 1-D electron-hole plasma with strong Coulomb correlations.

A laser structure containing 20 quantum wires with 14 nm $\times$ 6 nm lateral size was grown by the cleaved-edge overgrowth method with molecular beam epitaxy \cite{PfeifferAPL}. As schematically shown in Fig. 1, the active region of the laser consists of a 6 nm GaAs quantum well (arm well) with an Al${}_{0.5}$Ga${}_{0.5}$As top barrier and 20 periods of 14.15 nm Al${}_{0.073}$Ga${}_{0.927}$As quantum wells (stem wells) separated by 41.88 nm Al${}_{0.35}$Ga${}_{0.65}$As barriers. The 20 wire states (T-wires) are formed quantum-mechanically at the T-shaped intersections of the arm well and the stem wells \cite{SzymanskaPRB}.  The active region is itself embedded in a core of a T-shaped optical waveguide. A laser bar with 500 $\mu$m cavity length was cut from the wafer by cleavage, and the cleaved cavity-mirror surfaces were left uncoated. The growth method and the optical waveguide are similar to those in the previous work \cite{WegscheiderPRL,RubioSSC} except for additional annealing at 600 ${}^{\circ}$C for 10 minutes after the growth of the arm well, that has been found to dramatically improve the arm well flatness and uniformity \cite{AkiyamaSSC,YoshitaAPL}.  The high uniformity of the sample was characterized by micro-PL and PL-excitation (PLE) measurements under weak point excitation, which is reported elsewhere \cite{AkiyamaPRB} together with details of fabrication procedures. 

\begin{figure}
\includegraphics[width=.45\textwidth]{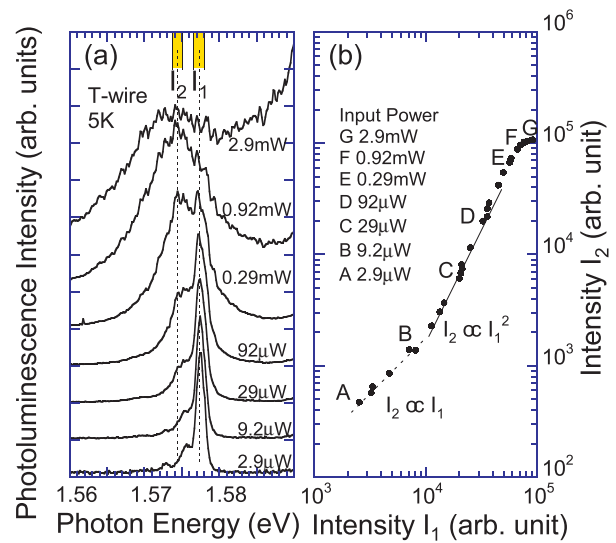}
\caption{
(a) PL spectra of T-wires at 5 K for various excitation powers at 1.634 eV measured via point spectroscopy on a spot of about 1$\mu$m diameter. (b) Plots of PL intensities $I_{1}$ and $I_{2}$ within the hatched energy windows in (a) for free excitons and a low-energy emission band, respectively. 
}
\label{2}
\end{figure}

\begin{figure}
\includegraphics[width=.35\textwidth]{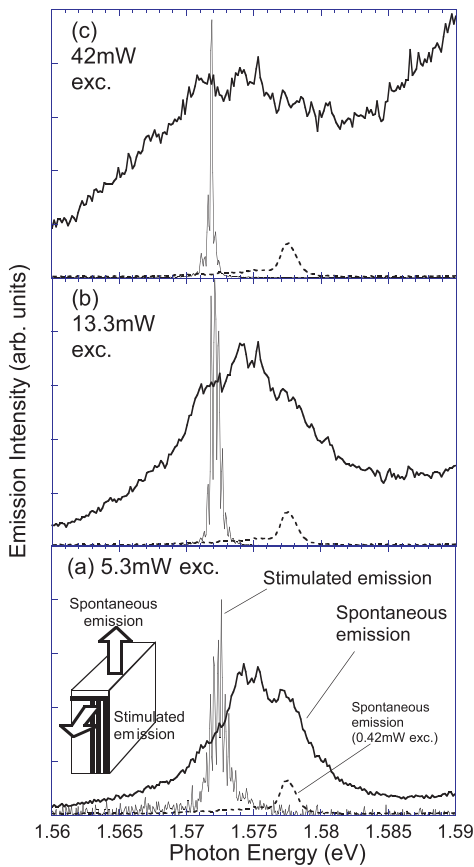}
\caption{
Stimulated emission spectra (thin solid curves) with simultaneously measured spontaneous emission spectra (thick solid curves) of a 500 $\mu$m-cavity laser with 20 wires at 5 K. The inset shows configuration of the measurements. It is optically pumped by 0.25 ms-pulses with 1\%-duty ratio at 1.634 eV. Indicated input power $I_{in}$ of (a) 5.3 mW, (b) 13.3 mW, and (c) 42 mW shows pulse height power incident on the laser. The spontaneous emission spectrum for $I_{in}$=0.42 mW is shown by a dashed curve to indicate an exciton PL peak. 
}
\label{3}
\end{figure}

\begin{figure}
\includegraphics[width=.4\textwidth]{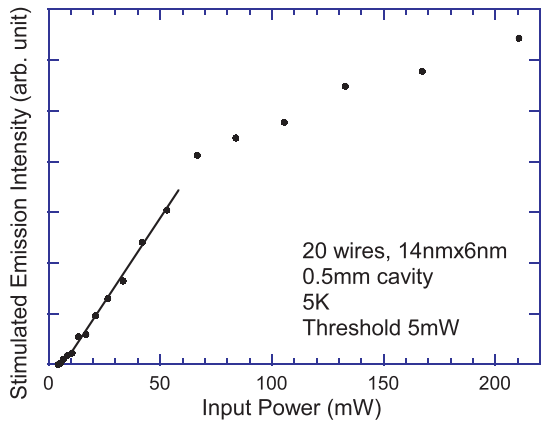}
\caption{
Stimulated emission intensity of a 500 $\mu$m-cavity laser with 20 wires at 5 K. It is optically pumped by 0.25 ms-pulses with 1\%-duty ratio at 1.634 eV. Indicated input power shows pulse height power incident on the laser. A straight line is drawn to guide the eyes. 
}
\label{4}
\end{figure}

All measurements here were done at 5 K, where the laser sample was attached to a copper block with silver paint, and mounted on the cold finger of a helium-flow cryostat. Output of a cw titanium-sapphire-laser was used in optical excitation at photon energy of 1.634 eV, which is resonant with the exciton absorption of the stem wells. Spontaneous emission spectra at various excitation powers were measured with point focus of about 1 $\mu$m. For stimulated emission measurements, two cylindrical lenses and a 0.4 numerical-aperture objective lens were used to focus the incident beam into a filament shape with about 1 $\mu$m width to pump the whole 500 $\mu$m-long laser cavity through the arm well surface (the top surface in Fig. 1), and the cw output of the excitation laser was mechanically chopped into 0.25 ms rectangular pulses of 1\% duty ratio to minimize sample heating. The peak input power per pulse, $I_{in}$, was varied from 0 to 210 mW. The stimulated emission was collected in the direction of the waveguide through one of the cavity-mirror surfaces (the front and rear surfaces in Fig. 1) and spontaneous emission was simultaneously measured in the direction perpendicular to the waveguide through the top surface of the arm well.

Figure 2 (a) shows normalized point-excitation PL spectra of the T-wires plotted for several excitation powers. The sharp PL peak of free excitons dominates the spectrum at the lower pump powers. The free exciton peak has narrow width of 1.5 meV and displays a small Stokes shift of 0.5 meV from the position of the free exciton absorption measured in PLE spectra, and is highly uniform over the whole wire length of 500 $\mu$m \cite{AkiyamaPRB}. Much weaker PL peaks in the low energy region vary in intensity at different positions in the sample, and are ascribed to localized excitons. As the pumping level is increased, a new emission band with maximum intensity $I_{2}$ grows at 3.2 meV below the free-exciton peak with intensity $I_{1}$. With increasing pump power this new emission band becomes broader but undergoes no further red-shift. At the higher pump levels the low-energy emission band dominates the spectrum and the free exciton peak is quenched.

Figure 2 (b) shows log plots of $I_{2}$ as function of $I_{1}$. We find that $I_{2}\propto I_{1}$ at the lowest pump powers due to contribution of localized excitons in $I_{2}$. For higher excitation levels we find that $I_{2}\propto I_{1}^2$. With increasing pump levels we first find larger exponents followed by saturation. In the region of $I_{2}\propto I_{1}^2$, biexcitons may contribute to the red-shifted PL band. For intense photo-excitation the free exciton peak is quenched and is not fully differentiated from the broadened lower energy PL band. In this regime the intense optical emission should be ascribed to a 1-D electron-hole plasma confined to the quantum wires. Here densities should be high enough so that there are no long-lived excitons or exciton complexes. Nevertheless, the low energetic position of the emission, and the absence of  further red-shift with photoexcitation at high pump levels shown in Fig. 2 (a) suggests that the {\it instantaneous} electron-hole correlations are strong. The state is unlike a free electron-hole plasma, and is better described as a neutral electron-hole plasma in which Coulomb correlations fix the peak emission energy to a value that is close to that of the biexciton energy.  We invoke Coulomb interactions to interpret the observation that emitted photon energies are lower but remain relatively close to the free exciton energy.

Figure 3 shows stimulated emission spectra (thin solid curves) of the T-wire laser for $I_{in}$ of 5.3 mW, 13.3 mW, and 42 mW. Spectral widths are here limited by the system resolution of 0.15 meV.  Lasing threshold is close to $I_{in}$=5.3 mW, where stimulated emission spectra show features of multi-mode lasing with about 10 longitudinal modes. As the input power is increased the number of modes decrease. For input powers of 30 mW single mode lasing is observed. The additional red-shift in laser emission from threshold at about $I_{in}$=5.3 mW is small, only 0.5 meV at $I_{in}$=42 mW, and 2 meV at $I_{in}$=210 mW. The observed behaviors are similar to those in previous results \cite{WegscheiderPRL,RubioSSC} except for the lower threshold and the smaller numbers of initial modes.  Figure 4 shows stimulated emission intensities plotted against input powers of up to 210 mW.  The greatly reduced threshold is a result of higher uniformity and smaller roughness than previous wire lasers \cite{WegscheiderPRL,RubioSSC}. This suggests that gain in T-wire lasers is not due to localized excitons. For high input power at around $I_{in}$=100 mW, the intensity of T-wire lasing saturates.  

Figure 3 also displays spontaneous emission spectra of the T-wire laser sample measured simultaneously with the stimulated emission spectra. In these measurements we employed the geometry shown in the inset to Fig. 3 (a). The dashed traces show the spontaneous emission spectrum for very weak excitation of $I_{in}$=0.42 mW. These spectra identify the location of the free exciton emission peak. The spontaneous emission spectra at high excitation levels in Figs. 3 (a) - 3 (c) are very similar to the PL results of Fig. 2 measured with point excitation.  

The results shown in Fig. 3 reveal that T-wire lasing is observed about 5 meV below the free exciton energy. Since there is no overlap between the lasing energy and the free exciton peak, gain for lasing cannot be due to free exciton recombination. Instead, the lasing photon energy overlaps the red-shifted broad PL band. We may conclude that, gain for lasing could be ascribed to the electron-hole plasma with strong Coulomb interactions among the photoexcited particles. The lasing energy is on the low energy side of the plasma emission band, presumably because some absorption may reduce gain near the peak of spontaneous emission.

The present results are qualitatively different from those in the first report of T-wire lasers \cite{WegscheiderPRL}, that were fabricated in structures with much larger roughness where the PL at low pump levels showed a broad band width of 10 meV caused by thickness fluctuations of up to several monolayers due to interface roughness. Thus, the PL band in Ref. \cite{WegscheiderPRL} could arise from localized excitons with Stokes shifts comparable to the PL linewidth. In such rougher quantum wires it may not have been possible to resolve a shift of the lasing emission from the free exciton peak.

It is interesting to point out that Fig. 3 (a) shows that near lasing threshold, for $I_{in}$=5.3 mW, the spontaneous emission spectrum while dominated by the broad PL band still shows a distinct free exciton peak. This is different from higher excitation regime, where the exciton peak is nearly completely quenched. Appearance of a distinct free exciton peak again points to the dominance of electron-hole Coulomb correlations over electron-electron or hole-hole Coulomb correlations \cite{RossiPRBTassonePRLDassarmaPRL,AkiyamaSSC}. It is thus conceivable that the Coulomb correlated system responsible for lasing gain in this regime could adequately be described as a multi-exciton complex. This is an intriguing issue that could be explored in further experiments. 

It should be stressed, however, that our results unambiguously exclude free excitons as the states that create gain in the T-wire lasers and show that stimulated emission is associated with gain due to a dense system of photo-excited electrons and holes confined in the quantum wires, which are strongly linked by Coulomb interactions. 

Finally, we highlight again that the lasing energy and the PL peak energy of the dense 1-D electron-hole system in photoexcited quantum wires stay at fixed energetic positions, showing almost no shift against pumping power. This behavior manifests strong internal electron-hole Coulomb correlations in 1-D system \cite{RossiPRBTassonePRLDassarmaPRL,AkiyamaSSC}. In 2-D and 3-D semiconductor lasers showing large lasing-energy shifts against input powers, gain is believed to come from recombination of a {\it free} electron-hole plasma, where a gain peak is at the band edge between free electrons and holes that red-shifts due to band-gap renormalization. 

In summary, in a highly uniform T-quantum-wire laser we achieved low threshold at 5 K. Spontaneous emission spectra show a sharp peak of free excitons at low excitation levels. A broader red-shifted PL band emerges at high pump power. This emission is identified as arising from the Coulomb-correlated electron-hole plasma with emission that overlaps the energies of biexcitons and higher multi-exciton states. The photon energies of the laser occur in range of plasma recombination and are red-shifted by 5 - 7 meV from the free exciton peak. Gain for lasing in highly uniform T-wires is ascribed to a strongly Coulomb-correlated electron-hole plasma instead of free or localized excitons.  

One of the authors (H. A.) acknowledges the financial support from the Ministry of Education, Culture, Sports, Science and Technology, Japan.

\clearpage

\end{document}